\documentclass[conference]{IEEEtran}

\usepackage{hhline}
\usepackage[numbers]{natbib}
\usepackage[utf8]{inputenc}
\usepackage{graphicx}

\usepackage{subfig}
\usepackage{multirow}
\usepackage{threeparttable}
\usepackage{eurosym,makecell,comment}
\usepackage{url}
\usepackage{xcolor}
\usepackage{pgfplots}
\usepackage{tikz}
\usepackage{booktabs}
\usepackage{float}

\usepackage{array}
\newcolumntype{P}[1]{>{\raggedright\arraybackslash}p{#1}}
\usepackage{dblfloatfix}

\DeclareRobustCommand*{\IEEEauthorrefmark}[1]{%
  \raisebox{0pt}[0pt][0pt]{\textsuperscript{\footnotesize #1}}%
}

\begin{document}

\title{A Lightweight Moving Target Defense Framework for Multi-purpose Malware Affecting IoT Devices}

\author{
    \IEEEauthorblockN{Jan von der Assen\IEEEauthorrefmark{1}, Alberto Huertas Celdr\'an\IEEEauthorrefmark{1}, Pedro M. S\'anchez S\'anchez\IEEEauthorrefmark{2}, Jordan Cedeño\IEEEauthorrefmark{1}, \\G\'er\^ome Bovet\IEEEauthorrefmark{3}, Gregorio Mart\'inez P\'erez\IEEEauthorrefmark{2}, Burkhard Stiller\IEEEauthorrefmark{1}}
    
    \IEEEauthorblockA{\IEEEauthorrefmark{1}Communication Systems Group CSG, Department of Informatics, University of Zurich UZH, CH--8050 Zürich, Switzerland \\{[vonderassen, huertas, cedeno, stiller]}@ifi.uzh.ch}
    \IEEEauthorblockA{\IEEEauthorrefmark{2}Department of Information and Communications Engineering, University of Murcia, 30100--Murcia, Spain \\{[pedromiguel.sanchez, gregorio]}@um.es}
    \IEEEauthorblockA{\IEEEauthorrefmark{3}Cyber-Defence Campus, armasuisse Science \& Technology, CH--3602 Thun, Switzerland \\gerome.bovet@armasuisse.ch}
}

\maketitle

\begin{abstract}

Malware affecting Internet of Things (IoT) devices is rapidly growing due to the relevance of this paradigm in real-world scenarios. Specialized literature has also detected a trend towards multi-purpose malware able to execute different malicious actions such as remote control, data leakage, encryption, or code hiding, among others. Protecting IoT devices against this kind of malware is challenging due to their well-known vulnerabilities and limitation in terms of CPU, memory, and storage. To improve it, the moving target defense (MTD) paradigm was proposed a decade ago and has shown promising results, but there is a lack of IoT MTD solutions dealing with multi-purpose malware. Thus, this work proposes four MTD mechanisms changing IoT devices' network, data, and runtime environment to mitigate multi-purpose malware. Furthermore, it presents a lightweight and IoT-oriented MTD framework to decide what, when, and how the MTD mechanisms are deployed. Finally, the efficiency and effectiveness of the framework and MTD mechanisms are evaluated in a real-world scenario with one IoT spectrum sensor affected by multi-purpose malware.

\end{abstract}


%

\IEEEpeerreviewmaketitle

\section{Introduction}

Society is experiencing a massive increment of Internet of Things (IoT) devices deployed over multiple real-world scenarios such as industry, home, health, transport, or agriculture. The IoT devices used in the previous environments present particularities in terms of hardware, operating systems, services, and data. However, there are also remarkable inter-scenario similarities such as the connectivity to the internet and limitations in terms of CPU, memory, and storage. These aspects, combined with the complexity of deploying detection and mitigation cybersecurity mechanisms, make IoT devices one of the most desired targets for cyberattacks~\cite{sanchez2021survey}.

Cybersecurity studies acknowledge this fact, since the number of cyberattacks affecting heterogeneous IoT devices is increasing every year~\cite{alsheikh2021state}. In addition, cybercriminals are moving towards the usage of multi-purpose malware, where malicious behaviors such as remote control, data leakage, encryption, mining, code execution hiding, and other hostile actions can be perpetrated by individual malware samples. This fact complicates the challenge of defending IoT devices since detection and mitigation mechanisms must be varied, and they usually consume many resources.

Assuming that perfect security is not realistic in any system, in 2009, a novel cyberdefense paradigm called Moving target defense (MTD) was introduced~\cite{cai2016moving}. MTD proposes to continuously change different aspects (like, for example, network, data, or runtime environment) of a given device or system to prevent or mitigate ongoing or future cyberattacks. Therefore, the idea is to reduce the attack surface and make it more difficult for attackers to exploit vulnerabilities.

Despite the progress achieved by existing MTD-based solutions, as stated in \cite{navas:2020:mtd}, a limited amount of work is dedicated to IoT devices. More in detail, the following challenges are still open: \textit{ch1)} lack of IoT MTD mechanisms moving different parameters to mitigate multipurpose malware affecting data integrity, availability, and confidentiality; \textit{ch2)} lack of on-host systems suitable for IoT devices and able to deploy MTD mechanisms proactively and reactively, and \textit{ch3)} lack of work evaluating the efficiency of MTDs in real testbeds. To improve these challenges, the main contributions of this work are:

\begin{itemize}
    \item Four MTD mechanisms changing the network, data, and runtime environment of IoT devices to mitigate multi-purpose malware (covering \textit{ch1} and code available in~\cite{github}).
    
    \item A lightweight and IoT-oriented MTD framework able to decide what, when, and how the proposed four MTDs are deployed in a reactive and proactive way (covering \textit{ch2}).
    
    \item The evaluation of the effectiveness and efficiency of the framework and the MTD mechanisms in a real-world scenario with a spectrum sensor affected by multi-purpose malware showing behaviors of Botnet, Ransomware, Rootkit, and Backdoor (covering \textit{ch3}).
\end{itemize}

The remainder of this article is organized as follows. Section~\ref{sec:related} reviews existing IoT MTD mechanisms.
While Section~\ref{sec:mtd} presents four novel IoT MTDs, Section~\ref{sec:framework} provides a framework to manage them. Section~\ref{sec:experiments} evaluates and compares the framework performance and efficiency in a real-world scenario. Finally, Section~\ref{sec:conclusion} draws conclusions and next steps.

\section{Related Work}
\label{sec:related}

A recent review of IoT MTD techniques concludes that IoT MTD is still immature, and novel techniques deployed in real-world scenarios should be prioritized~\cite{navas:2020:mtd}. Moreover, the authors present \textit{WHAT}, \textit{WHEN}, and \textit{HOW} as design principles for MTDs. The WHAT represents the components of the system that are changed to secure the system, being \textit{Network}, \textit{Data}, \textit{Software}, \textit{Runtime Environment}, and \textit{Platform} the most representative. The WHEN indicates the moment in which the system should change the WHAT, and The HOW is the way in which the WHAT is moved. Finally, the authors categorized the 32 existing IoT MTD works according to the WHAT principle, where 54\% focused on the network, 20\% on runtime, 13\% on software, 10\% on data, and 3\% on the platform.

Starting from the network category, the one with the most MTDs, \cite{zeitz:2018:net} proposes \textit{$\mu$MT6D}, an MTD mechanism oriented to constrained devices that limit the time window for reconnaissance attacks through IP address rotation. 
\cite{nizzi:2019:ip} presents AShA, a method allowing a fast, secure, and collision-free address renewal of IPv6. The previous two solutions have been tested in simulated environments before the malware infection happens. In contrast, the work at hand not only protects against network attacks (proactively and reactively) but also against attacks affecting data and runtime environments. Continuing with the runtime category, \cite{Habibi:2015:runtime} proposes the combination of software and hardware mitigation. From the software perspective, it uses a randomization approach that modifies the layout of the executable code, preventing code-reuse attacks. From hardware, it isolates the binary code to avoid exposing executed code to attackers. Compared to the work at hand, this solution fails against attacks manipulating links to libraries. 
Dealing with the data category, \cite{vuppala:2020:data} presents a side-channel resilient MTD mechanism for electromagnetic-based side-channel attacks. The proposed solution applies rekeying at suitable intervals to reduce the computational and communication overhead. Another MTD focused on data is \cite{puthal:2016:dlsef}. It proposes a Dynamic Key-Length-Based Security Framework (DLSeF) that uses a shared key that is frequently updated to ensure end-to-end security. Compared to the work at hand, the previous two solutions do not prevent attacks affecting data encryption.

\begin{table}[ht!]
\centering
\caption{Comparison of IoT MTD Mechanisms. Network (N), Runtime (R), Data (D), Proactive (Pr), Reactive (Re).}
\label{tab:related}
\begin{tabular}{P{1.2 cm} P{1.5 cm} P{0.5 cm} P{1.9 cm} P{0.6 cm} P{0.5 cm}}

\textbf{Sol. Year} & \textbf{Attack} & \textbf{What} & \textbf{How} & \textbf{When} & \textbf{Eval.}\\
\hline \hline

\cite{zeitz:2018:net} 2018 & C\&C & N & IP Shuffle & Pr & Sim. \\\hline

\cite{nizzi:2019:ip} 2019 & C\&C & N & IPv6 Shuffle & Pr & Sim.\\\hline

\cite{Habibi:2015:runtime} 2015 & Code Reuse & R & Code Random. & Pr & HW \\\hline

\cite{vuppala:2020:data} 2020 & Side-channel & D & Encryption Key & Pr & - \\\hline

\cite{puthal:2016:dlsef} 2016 & Side-channel & D & App. Encr. Key & Pr & Sim. \\\hline

Ours & C\&C, Ransomware, Rootkit, Backdoor & N, R, D & IP \& Format Shuffle, Library Sanitation, Data Honeypot & Pr, Re & HW \\\hline

\end{tabular}
\end{table}

\tablename~\ref{tab:related} compares the main aspects covered by previous works. As can be seen, attacks performed by recent malware able to encrypt and steal sensitive data, change libraries, and hide themselves are not considered by existing MTDs. Furthermore, most MTDs are evaluated in simulated scenarios. Finally, there is no framework able to deploy MTD mechanisms when needed.

\section{IoT MTD mechanisms}
\label{sec:mtd}

This section presents novel MTD mechanisms to mitigate multi-purpose malware controlling IoT devices, encrypting data, hiding malicious commands, or leaking sensitive data.

\subsection{File Encryption}
\label{sub:encr}

This novel reactive MTD mechanism creates a honeypot with dummy files on the IoT device (victim). It traps ransomware in a dynamic expanding and collapsing directory and file tree with dummy files, in the meanwhile the encrypting process is discovered and killed. The goal is to keep as much data safe as possible, while focusing exclusively on the encryption behavior shared by multiple ransomware implementation. Related malicious phases commonly found in Ransomware, such as cryptographic key exchange with a C\&C server, are covered by other MTD techniques.

\begin{figure}[ht!]
    \centering
    \includegraphics[width=0.8\columnwidth]{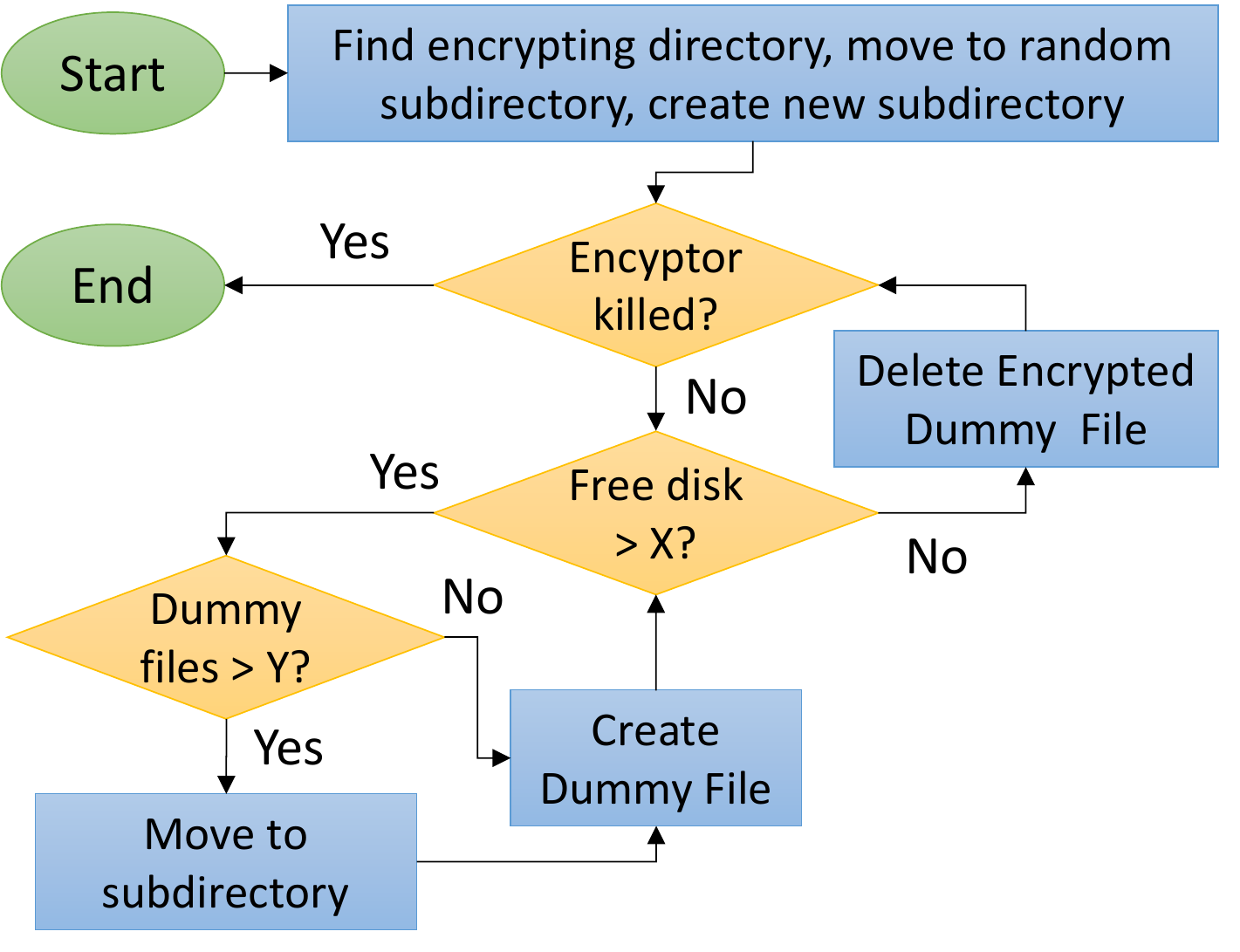}
    \caption{File Encryption Flowchart}
    \label{fig:flow}
\end{figure}

Before providing the MTD details, it is important to know that most of the existing ransomware samples recursively encrypt files. 
To mitigate it, first, the proposed MTD is deployed in the directory in which files are being encrypted. Then, it moves to a random existing subdirectory, and creates a new subdirectory with dummy files. Once a given number of dummy files per directory is created, it moves to a new subdirectory and repeats the process. This movement is done because once the encryption on a directory starts, no new files added to that directory are encrypted. Additionally, the MTD deletes dummy files when they are encrypted to avoid depleting scarce disk space. \figurename~\ref{fig:flow} shows a flowchart with the previous steps. This creation of dummy files is key to create a file-based honeypot that allows the ransomware to be stalled and identified. Lastly, the MTD monitors all running processes to identify and kill the encrypting process. For that, the CPU usage of all processes is monitored, and all processes falling below a minimum threshold are discarded due to encrypting files is a CPU-intensive task. In the next step, processes included in a whitelist are filtered out. Finally, the MTD monitors how many files are opened by each process of the suspicious list within a minute. If the number of open files exceeds a configurable threshold, the process is killed.

\subsection{File Format}
\label{sub:file}

This novel reactive MTD shuffles the extensions of IoT sensors files to hide them from malware such as backdoors and ransomware affecting data availability, integrity, or confidentiality. This MTD relies on the fact that some malware families select target files according to their extensions.

To achieve the previous functionality, this MTD creates pseudo extensions consisting of alphanumeric strings randomly generated and replaces selected file extensions with the pseudo ones. The MTD maintains a dictionary to track the relationship between valid and pseudo extensions. When a new pseudo extension is created, the MTD checks in the dictionary if it is valid or used. This is done to avoid collisions during reconstruction. Then, once malware is mitigated, the pseudo extensions are replaced with genuine ones. 

\subsection{Libraries}
\label{sub:libraries}

This novel MTD changes the Linux runtime environment of the IoT sensor reactively or proactively to reduce or mitigate manipulations of legitimate libraries to execute malicious code (done by rootkits). More in detail, this MTD sanitizes corrupted libraries and unlinks fake libraries.

When running applications on Linux, it is possible to preload needed libraries through the \textit{LD\_PRELOAD} environment variable contained in the \textit{/etc/ld.so.preload} file. In particular, LD\_PRELOAD links shared libraries that are preloaded on the user-space, taking precedence over preloaded libraries of the kernel-space. Therefore, malware like rootkits take advantage of this and modify this environmental variable to preload malicious libraries hiding themselves. Additionally, rootkits can unlink the \textit{/etc/ld.so.preload} file from the dynamic linker, and a malicious file is linked in its place. To prevent or mitigate these behaviors, this MTD sanitizes \textit{i)} the \textit{ld.so.preload} file with a backup containing the right value of the \textit{LD\_PRELOAD} variable, and \textit{ii)} the dynamic link to the \textit{/etc/ld.so.preload} file. Regarding the first action, the MTD guarantees that \textit{LD\_PRELOAD} points to the legitimate \textit{libc.so.6} library. Therefore, typical malicious behaviors of rootkits such as files hiding or disabled access files are prevented. Dealing with the second action, the MTD overwrites the \textit{/lib/arm-linux-gnueabihf/ld-2.24.so} file with the string \textit{/etc/ld.so.preload} to link again with the proper library.

\subsection{IP Address}
\label{sub:commMTD}

This is an adaptation of existing MTD mechanisms shuffling the IP address of IoT devices (victim) to mitigate the impact of malware based on C\&C (like Botnets). The main difference compared to related work is that it works with private addresses and can be executed proactively or reactively to disrupt the communication between the IoT sensor and C\&C. The MTD is designed so that it cuts communication to all types of C\&C servers, whether the server is running on the device or on an external host (locally or publicly reachable).

From the implementation perspective, the MTD mechanism creates a list with all IP addresses available in the private network where the IoT device is connected. It is done by generating a list with all IP addresses of the private network and then removing the IPs assigned to active devices (obtained with the \textit{arp-scan} command). After that, a given IP address from the list is randomly chosen, and the IoT device requests to migrate to that IP using \textit{ifconfig}. Then, the MTD checks if there is internet connectivity. If so, the migration is successful, and the legitimate services of the IoT device are restarted. If there is no connectivity, the MTD removes the IP from the list of possible ones, chooses a new IP, and checks its connectivity. This sequence is repeated until a successful migration.

\section{Lightweight Hybrid MTD Framework}
\label{sec:framework}

\begin{figure}[htpb!]
    \centering
    \includegraphics[width=0.7\columnwidth]{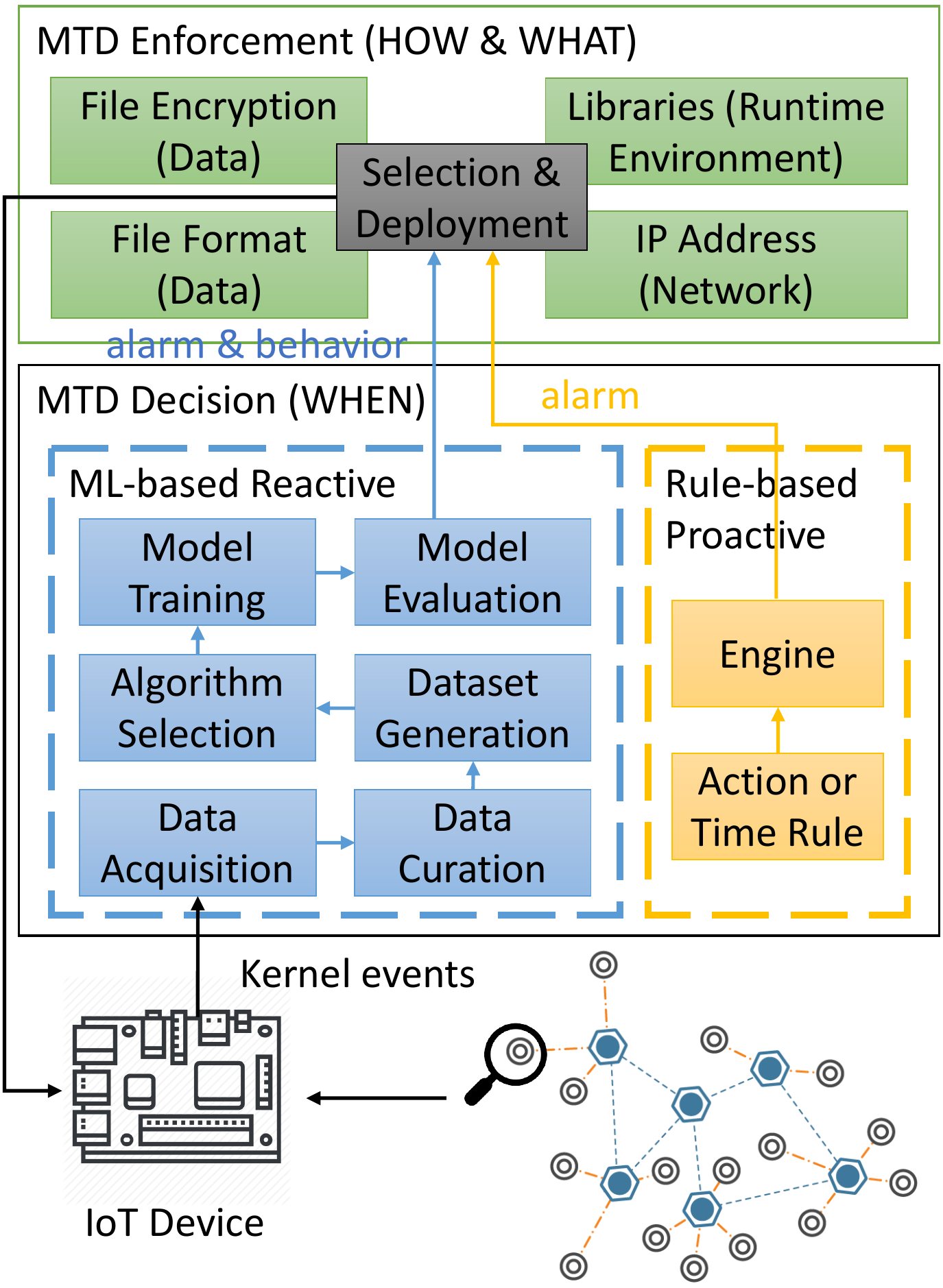}
    \caption{IoT-oriented MTD Framework Architecture}
    \label{fig:architecture}
\end{figure}

This section presents a novel MTD framework for IoT devices. The framework objective is to decide WHEN, HOW, and WHAT MTD mechanisms explained in Section~\ref{sec:mtd} should be deployed to mitigate multi-purpose malware. \figurename~\ref{fig:architecture} shows the modules and components of the proposed framework, which are all deployed on the IoT device.

\begin{itemize}
    \item \textit{MTD Decision.} This module decides \textit{WHEN} to deploy MTD mechanisms. 
    
    \item \textit{MTD Enforcement.} This module deals with HOW and WHAT MTD mechanisms are deployed.
    
\end{itemize}

The MTD Decision module focuses on WHEN to deploy MTD mechanisms according to proactive and reactive criteria. For the proactive approach, the system administrator defines in the \textit{Action or Time Rule} component the actions or time windows that need to be met by the IoT device to deploy MTDs. Periodically, the \textit{Engine} component checks the criteria and sends an alarm to the MTD Enforcement module. This approach is the most lightweight and does not need previous knowledge to make the WHEN decision. However, the lack of information also complicates further decisions about HOW and WHAT MTD should be deployed. 

The reactive approach can facilitate this decision, but it increases the computational complexity and decision time of the MTD Decision module. In particular, the reactive process decides WHEN to deploy MTDs according to the output of a supervised Machine Learning (ML)-based process able to classify the IoT device behavior. More in detail, first, the \textit{Data Acquisition} component periodically collects kernel events of the IoT device related to CPU, memory, file system, network interface, scheduler, drivers, and random numbers usage. Then, the \textit{Data Curation} component preprocesses the events and extracts relevant features. Feature vectors are stored and labeled in a dataset created by the \textit{Dataset Generation} component. The monitoring, data curation, and dataset generation processes are repeated for normal and malicious behaviors of multi-purpose malware. After that, the \textit{Algorithm Selection} component selects a set of classification algorithms, the \textit{Model Training} trains the ML models with the created dataset, and the \textit{Model Evaluation} detects the IoT device behavior, which is sent to the upper module.

The MTD Enforcement module decides HOW and WHAT MTD mechanism (proposed in Section~\ref{sec:mtd}) should be deployed according to the outputs of the previous module. For that, the \textit{Selection \& Deployment} component defines a set of rules deciding the MTD mechanisms that should be deployed. If the IoT device behavior is available (reactive approach), it deploys the most suitable MTD for that behavior. If not (proactive approach), all proactive MTDs are deployed simultaneously. An example of policies is provided below.

\begin{itemize}
    \item If [Rootkit] $\rightarrow$ deploy [Libraries MTD]
    \item If [Ransomware] $\rightarrow$ deploy [File Encryption MTD]
    \item If [Botnet] $\rightarrow$ deploy [IP Address MTD]
    \item If [Backdoor] $\rightarrow$ deploy [File Format MTD]
    \item If [action] or [time] $\rightarrow$
    deploy [File Format MTD] and [IP Address MTD] and [Libraries MTD] 
\end{itemize}

\section{MTD Framework Efficiency and Effectiveness}
\label{sec:experiments}

It describes the implementation and the experimental results achieved by the MTD framework when deployed on a real IoT spectrum sensor affected by multi-purpose malware. 

\subsection{IoT Spectrum Sensors Affected by Multi-purpose Malware}

The IoT evolution has brought Radio Frequency crowdsensing platforms to reality, offering the optimization of spectrum usage. However, sensors used by these platforms are resource-constrained devices with well-known vulnerabilities and limitations to deploy complex cybersecurity mechanisms. Attackers are aware of this fact, and the increment of multi-malware affecting IoT devices proves it.

In the context of malware affecting IoT devices, this work has considered ElectroSense~\cite{electrosense}, a real-world, open-source, and crowdsensing solution that uses Raspberry Pis to continuously send RF spectrum data gathered from connected software-defined radio kits. In particular, this work has deployed an ElectroSense sensor in a Raspberry Pi 4 with 1.5 GHz CPU and 3.7 GB RAM. The Raspberry Pi 4 executes the official and publicly available ElectroSense software, and is connected to a local area network with access to the Internet. Then, the device has been infected with multi-purpose malware such as Bashlite (botnet); Ransomware\_PoC (ransomware); Beurk and Bdvil (rootkit); and TheTick, PythonBackdoor and httpBackdoor (backdoor). All these malware samples and some others can be found and downloaded through \cite{celdran2022intelligent}. Furthermore, since the main contribution of this work focuses on MTD techniques and not malware detection, other malware samples detectable by the proposed solution can be found in the previous work. These malware samples can be executed by exploiting weak passwords or well-known vulnerabilities of IoT devices network services such as SSH or Telnet. Furthermore, their functionality is to control sensors remotely, encrypt data, hide malicious actions, leak sensitive data, or execute malicious code. Finally, the impact of these behaviors on spectrum sensors is to \textit{i)} disrupt crowdsensing platforms services, \textit{ii)} execute Distributed-Denial of Service (DDoS) attacks, or \textit{iii)} perform lateral movement attacks.

\subsection{MTD Decision Module: ML-based Reactive}
\label{sub:detection}

For optimal deployment of MTD mechanisms, detecting the malicious behavior of IoT spectrum sensors is needed. In this sense, the Data Acquisition component uses kernel software events, present in any Linux system due to their flexibility and variety. To monitor this dimension, \textit{perf} Linux tool is selected, monitoring $\approx$80 values from different event families such as network, memory, file system, CPU, process scheduler, or device drivers. Every 5 s, The monitoring collects the data of the previous metrics, considering all processes running in the sensor. Then an extra processing time is required for \textit{perf} to calculate and return the values. \tablename~\ref{tab:monitoring_resources} shows the resource usage of the Data Acquisition component. As it can be appreciated, it implies a low resource consumption, with only a maximum 4\% of usage in one CPU core. Besides, the complete \textit{perf} monitoring and data processing loop is performed in $\approx$10 s, which is an acceptable monitoring time for early malware detection.

\begin{table}[htpb]
	\caption{Behavior Monitoring Resource Consumption.}
	\centering
    \begin{tabular}{m{1.4cm}m{1.4cm}m{1.3cm}m{1.3cm}m{1.3cm}}
    
        \makecell[c]{\textbf{Monitoring}\\\textbf{Time}} & \makecell[c]{\textbf{Processing}\\\textbf{Time}} & \makecell[c]{\textbf{CPU}\\\textbf{Usage}} & \makecell[c]{\textbf{Memory}\\\textbf{Usage}} & \makecell[c]{\textbf{Storage}\\\textbf{Usage}} \\
        
        \hline \hline

        \makecell[c]{5 secs} & \makecell[c]{$\approx$5.21 secs} & \makecell[c]{1-4 \%} & \makecell[c]{6.14 MB} & \makecell[c]{6.98 kB} \\
        \hline
    \end{tabular}
    \label{tab:monitoring_resources}
\end{table}{}

After that, since the framework uses supervised ML/DL algorithms, it is necessary to monitor each behavior to allow the algorithms to be trained and generate models capable of differentiating malicious activities. Therefore, normal (uninfected) behavior and the different malware samples (Beurk, Bdvl, Bashlite, Ransomware\_PoC, HttpBackdoor, PythonBackdoor, and TheTick) are monitored for a minimum of six hours. In total, $\approx$2160 vectors are available per behavior, making up a dataset of $\approx$21600 vectors and $\approx$6.78 MB.

For attack identification, k-NearestNeighbors (k-NN), Support Vector Machine (SVM), XGBoost (XGB), Decision Tree (DT), Random Forest (RF), and Multi-Layer Perceptron (MLP) are tested. Then, the available dataset is divided into 80\% for training (including cross-validation) and 20\% for testing. Besides, data standardization has been applied using \textit{min-max} for the algorithms that require normalization. \tablename~\ref{tab:clasif_alg_hyp} shows the classification results per model in terms of average F1-Score. RF and XGB are the models providing the best performance with a 0.98 average F1-Score. For deployment, RF is selected due to its shorter training time and lower memory and storage usage. In this sense, the training in the RPi4 takes $\approx$39.46 secs (using one CPU core at 100\%), while the preprocessing and evaluation of a single vector takes $\approx$0.0427 s. Besides, the model needs 89.55 MB in memory and storage, since \textit{pickle} library is used to serialize the binary object from memory to a file.

\begin{table}[htpb!]
	\caption{F1-Score per Classification Algorithm.}
	\centering
    \begin{tabular}{m{1.8cm}m{0.7cm}m{0.65cm}m{0.65cm}m{0.65cm}m{0.65cm}m{0.65cm}}
         
         \makecell[c]{\textbf{Algorithm}} & k-NN & SVM & XGB & DT & RF & MLP\\
         \hline \hline
         \makecell[c]{\textbf{Avg.} \textbf{F1-Score}} & 0.88 & 0.94 & 0.98 & 0.97 & 0.98 & 0.95 \\
         \hline
         
    \end{tabular}
    \label{tab:clasif_alg_hyp}
\end{table}{}

\begin{figure}[htpb!]
    \centering
    \includegraphics[width=1\columnwidth,trim={0 200 60 0} ,clip=true]{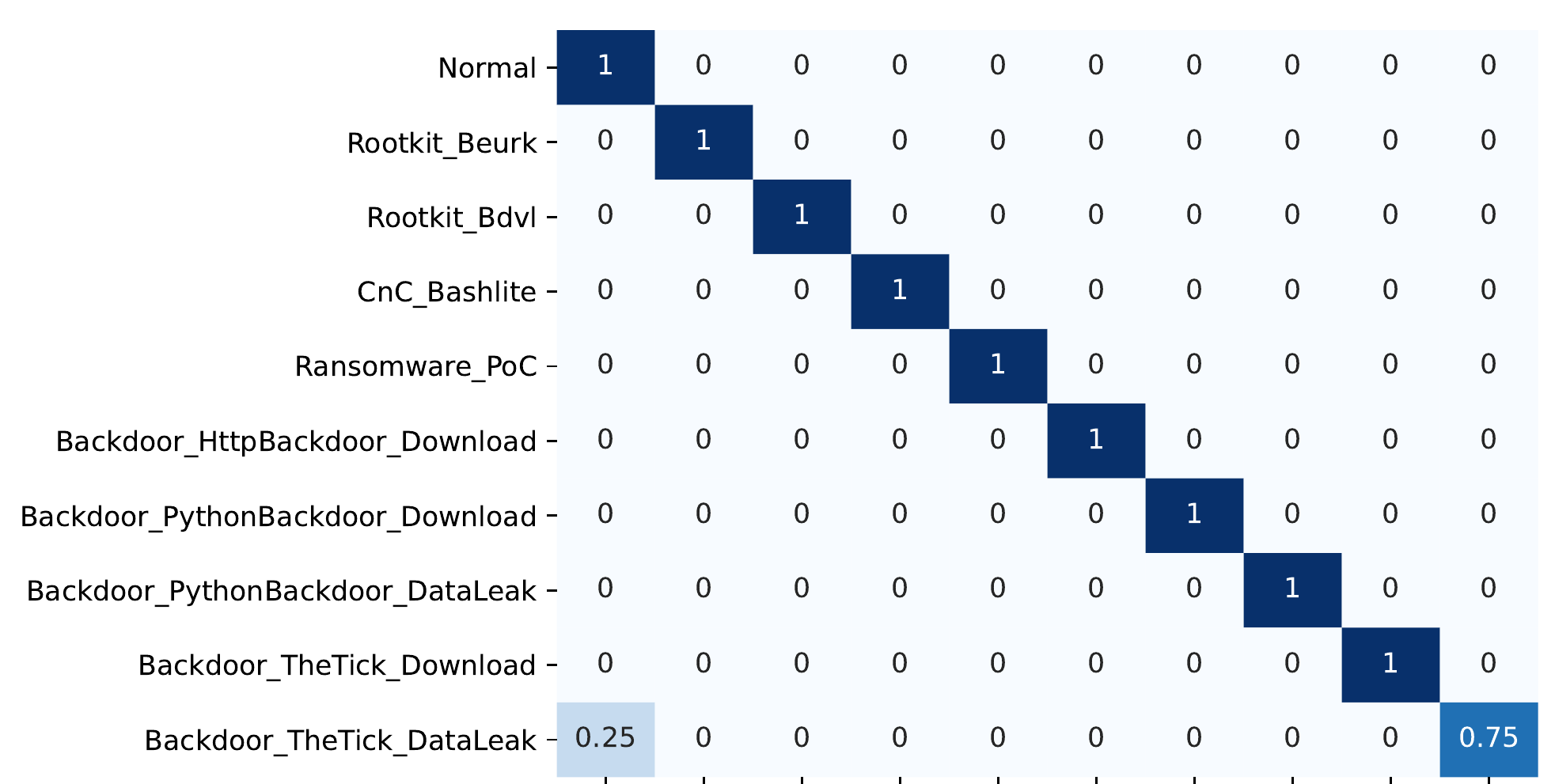}
    \caption{RF Confusion Matrix}
    \label{fig:matrix}
\end{figure}

Finally, \figurename~\ref{fig:matrix} shows the confusion matrix for the classification of each behavior and malware. It can be seen that almost all behaviors are correctly classified, only having trouble detecting the 25\% of the times the Data Leak attack when using TheTick backdoor.

In conclusion, the ML-based Reactive component takes $\approx$10 s to accurately detect normal and malicious behaviors when deployed as part of the MTD framework (only a few false positives for Data Leak are present). In addition, the consumption of CPU, RAM, and storage during monitoring and model training/evaluation is acceptable for IoT spectrum sensors implemented in Raspberry Pis.

\subsection{MTD Enforcement Module: Deploying MTD Mechanisms}

A set of experiments has been conducted to analyze the efficiency and effectiveness of the framework and the four MTD mechanisms. On the one hand, to evaluate the reactive deployment of MTD mechanisms, each experiment: \textit{i)} infects the IoT spectrum sensor with the appropriate malware, \textit{ii)} detects the malware and malicious behavior, and \textit{iii)} deploys the MTD mechanism. On the other hand, the MTD deployment relies on time intervals for proactive evaluation.

\tablename~\ref{tab:mtd_performance} shows the duration of each phase (highlighting the runtime of the respective malicious behavior or MTD technique), the CPU, RAM, time, and I/O blocks/s used in the sensor. In addition, it shows the KB/s sent by the spectrum scanning process. Finally, these metrics are measured for \textit{i)} the spectrum sensor (with and without MTD framework), and \textit{ii)} the MTDs deployed reactively and proactively.


\begin{table}[htpb!]
	\caption{Effectiveness and Efficiency of MTD Enforcement}
	\centering
    \begin{tabular}{m{2.3cm} m{0.9cm}m{0.6cm}m{0.6cm}m{0.9cm}m{0.9cm}}
         
         \makecell[l]{\textbf{Scenario}}   & \textbf{Duration}  & \textbf{CPU}     & \textbf{RAM}       &   \textbf{I/O}        & \textbf{Spectrum}\\
         \makecell[l]{\textbf{}}           &    \textbf{[s]}    &  \textbf{[\%]}   &   \textbf{[\%]}    &   \textbf{[Blocks/s]} & \textbf{[KB/s]} \\
         \hline \hline 

         \makecell[l]{\textbf{Sensor}} & 1035 & 23.37   & 15.53    & 0/0      & 16.34 \\
         \makecell[l]{\textbf{MTD Framework}} & 1035  & 23.94   & 15.99   & 0/0.3      & 15.86 
         \\
         \hline  \hline
         \makecell[l]{\textbf{\textit{Reactive MTD}}}
         \\\cmidrule{1-1}
         \makecell[l]{\textbf{Botnet}}    &  42  & 22.75   & 8.44     & 0/1.5        & 16.31     \\
         \makecell[l]{\textbf{C\&C MTD}}   & 12   & 19      & 8.43     & 0/6        & 15.10     \\
         \hline 
         \makecell[l]{\textbf{Data Leak}}  &  112   &  29.33 & 14.27  &   2/1        & 15.28     \\ 
         \makecell[l]{\textbf{File Format MTD}} &   56  & 35.20  & 14.29   &  3.60/1.50     & 15.14 \\
         \hline 
         \makecell[l]{\textbf{Backdoor}}   &  112  &  22    & 17.27   &  0/0.25    & 15.85     \\
         \makecell[l]{\textbf{C\&C MTD}}    &   14  &  21.5  & 17.46   &  0/1       &  9.37     \\
         \hline 
         \makecell[l]{\textbf{Ransomware}}   &  84   &  83.86    & 9.85   &  79/572  & 15.50 \\
         \makecell[l]{\textbf{Encryption MTD}}  & 70     &  90.66  & 9.89  & 93/667       &  15.64 \\
         \hline 
         \makecell[l]{\textbf{Rootkit}}    &  42   &  27.50 & 13.43    &  0/0.5        &   15.40      \\
         \makecell[l]{\textbf{Libraries MTD}} & 12   & 34.00  & 13.45   & 0/2         &   15.30 \\
         \hline \hline
         \makecell[l]{\textbf{\textit{Proactive MTD}}}
         \\\cmidrule{1-1}
         \makecell[l]{\textbf{Rootkit}}   &  29   &  40 & 11.71    &  0/25        &   16.51      \\
         \makecell[l]{\textbf{Libraries MTD}} & 1035   & 25.82  & 11.63   & 0/5.49         &   16.26 \\
         \hline
         
    \end{tabular}
    \label{tab:mtd_performance}
\end{table}{}

As can be seen, the MTD framework resource consumption in the background is minimum. For proactive and reactive deployment, the MTD mechanisms differ in mitigation time. For example, data leakage behavior impacts the device during 112 s. Upon deploying the \textit{File Format} MTD which operated for 56 s, critical files are protected so that only 11 MB of data were leaked out of 300 MB of PDF files. Furthermore, the CPU, RAM, I/O, and data collected by the sensor of the phase where the malware is running can be compared to the row underneath showing the impact of malware and MTD operation. From that, it can be concluded that for all MTD techniques, the resource impact is minimal when deployed reactively. From the proactive deployment perspective, the Libraries MTD does not impact the sensing service. Thus, the MTD is deployed every minute, leading to successful disinfection of rootkits and backdoors after 29 s with negligible impact on resources.

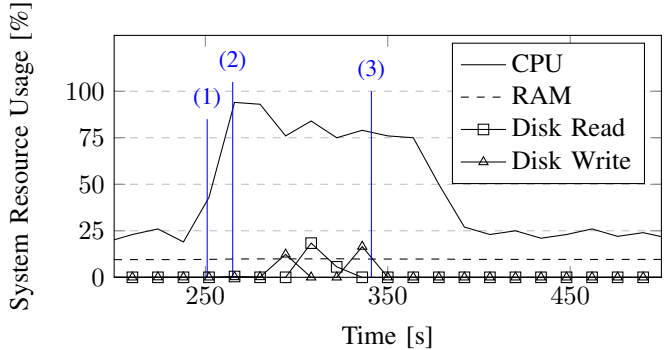
\begin{figure}[htpb!]
\begin{tikzpicture}
\begin{axis}[
    xlabel={Time [s]},
    ylabel={System Resource Usage [\%]},
    xmin=200, xmax=500,
    ymin=0, ymax=130,
    xtick={250, 350, 450},
    ytick={0, 25, 50, 75, 100},
    legend pos=north east,
    width=\linewidth,
    legend cell align={left},
    height=4.8cm,
    ymajorgrids=true,
    grid style=dashed,
]

\addplot[
    ]
    coordinates {
    (0,25)(14,17)(28,25)(42,23)(56,21)(70,21)(84,18)(98,23)(112,23)(126,25)(140,14)(154,18)(168,25)(182,19)(196,19)(210,23)(224,26)(238,19)(252,43)(266,94)(280,93)(294,76)(308,84)(322,75)(336,79)(350,76)(364,75)(378,50)(392,27)(406,23)(420,25)(434,21)(448,23)(462,26)(476,22)(490,24)(504,21)(518,21)(532,23)(546,25)(560,24)(574,21)(588,23)(602,24)(616,28)(630,20)(644,16)(658,22)(672,28)(686,16)(700,17)(714,24)(728,22)(742,20)(756,24)(770,25)(784,24)(798,17)(812,22)(826,23)(840,29)(854,21)(868,20)(882,26)(896,19)(910,25)(924,29)(938,26)(952,24)(966,27)(980,25)(994,19)(1008,18)(1022,23)(1036,19)
    };
    \addlegendentry{CPU}
    
\addplot[
    legend style={at={(0.02,0.02)},anchor=south west},
    color=black,
    dashed
    ]
    coordinates {
    (0,9.37303899310746)(14,9.38946681019749)(28,9.37099827048758)(42,9.38446703977879)(56,9.51599161263003)(70,9.45364753659271)(84,9.44334188736232)(98,9.44436224867225)(112,9.44436224867225)(126,9.45395364498569)(140,9.45130070557985)(154,9.44650500742313)(168,9.44956609135295)(182,9.45558622308159)(196,9.46997331755175)(210,9.5131346009622)(224,9.51395089001015)(238,9.53027667096919)(252,9.62986393481932)(266,9.84036447305991)(280,9.89556601992766)(294,9.89668841736859)(308,9.91556510160248)(322,9.96199154120474)(336,9.81689616293129)(350,9.81618191001434)(364,9.81546765709738)(378,9.80720273048687)(392,9.63588406654796)(406,9.62241529725676)(420,9.62302751404272)(434,9.63231280196318)(448,9.62996597095031)(462,9.59200853022055)(476,9.62578248957956)(490,9.62221122499477)(504,9.60772209439363)(518,9.61904810493396)(532,9.62282344178073)(546,9.6165992377901)(560,9.61966032171992)(574,9.62568045344857)(588,9.61221168415736)(602,9.61884403267197)(616,9.6115994673714)(630,9.61843588814799)(644,9.61251779255034)(658,9.62149697207781)(672,9.61731349070706)(686,9.60496711885679)(700,9.59211056635155)(714,9.59374314444745)(728,9.59935513165212)(742,9.58343749521706)(756,9.5910902050416)(770,9.61476258743221)(784,9.60068160135504)(798,9.60047752909305)(812,9.61976235785092)(826,9.60588544403573)(840,9.60792616665561)(854,9.60088567361703)(868,9.5996612400451)(882,9.61976235785092)(896,9.60690580534567)(910,9.60751802213164)(924,9.60915060022754)(938,9.60353861302287)(952,9.60119178201001)(966,9.60945670862052)(980,9.58935559081471)(994,9.5997632761761)(1008,9.5860904346229)(1022,9.61904810493396)(1036,9.61068114219245)
    };
    \addlegendentry{RAM}
    
\addplot[
    legend style={at={(0.02,0.02)},anchor=south west},
    color=black,
    mark=square
    ]
    coordinates {
    (0,0)(14,0)(28,0)(42,0)(56,0)(70,0.17574692442882248)(84,0)(98,0)(112,0)(126,0)(140,0)(154,0)(168,0)(182,0)(196,0)(210,0)(224,0)(238,0)(252,0)(266,0.4393673110720563)(280,0)(294,0)(308,18.36555360281195)(322,5.623901581722319)(336,0)(350,0)(364,0)(378,0)(392,0)(406,0)(420,0)(434,0)(448,0)(462,0)(476,0)(490,0)(504,0)(518,0)(532,0)(546,0)(560,0)(574,0)(588,0)(602,0)(616,0)(630,0)(644,0)(658,0)(672,0)(686,0)(700,0)(714,0)(728,0)(742,0)(756,0)(770,0)(784,0)(798,0)(812,0)(826,0)(840,0)(854,0)(868,0)(882,0)(896,0)(910,0)(924,0)(938,0)(952,0)(966,0)(980,0)(994,0)(1008,0)(1022,0)(1036,0)(1050,0)
    };
    \addlegendentry{Disk Read}
    
\addplot[
    legend style={at={(0.02,0.02)},anchor=south west},
    color=black,
    mark=triangle,
    ]
    coordinates {
    (0,0.4959159859976663)(14,0)(28,0.10210035005834306)(42,0.05834305717619603)(56,0.1750291715285881)(70,0)(84,0)(98,0.08751458576429405)(112,0)(126,0)(140,0)(154,0)(168,0)(182,0)(196,0)(210,0)(224,0.08751458576429405)(238,0.043757292882147025)(252,0)(266,0.1750291715285881)(280,0.1896149358226371)(294,12.310385064177364)(308,0)(322,0)(336,16.525670945157525)(350,0.043757292882147025)(364,0)(378,0)(392,0)(406,0)(420,0)(434,0)(448,0)(462,0)(476,0)(490,0)(504,0)(518,0)(532,0)(546,0)(560,0)(574,0)(588,0)(602,0)(616,0)(630,0)(644,0)(658,0)(672,0)(686,0)(700,0)(714,0)(728,0)(742,0)(756,0)(770,0)(784,0)(798,0)(812,0)(826,0)(840,0)(854,0)(868,0)(882,0)(896,0)(910,0)(924,0)(938,0)(952,0)(966,0)(980,0)(994,0)(1008,0)(1022,0)(1036,0)(1050,0)
    };
    \addlegendentry{Disk Write}
    
   \draw[blue] (axis cs: 251,0) -- 
       (axis cs: 251,85)node[anchor=center, above=0.5pt]{\small{(1)}};
   \draw[blue] (axis cs: 265,0) -- 
       (axis cs: 265,105)node[anchor=center, above=0.5pt]{\small{(2)}};
   \draw[blue] (axis cs: 341,0) -- 
       (axis cs: 341,100)node[anchor=center, above=0.5pt]{\small{(3)}};

\end{axis}
\end{tikzpicture}
\caption{Device Resource Impact When Mitigating Ransomware}
\label{fig:ransomware}
\end{figure}

In terms of CPU consumption, Ransomware, and the File Encryption MTD have the strongest impact as presented in \figurename~\ref{fig:ransomware}. Here, the three stages consisting of \textit{(1)} encryption, \textit{(2)} MTD operation, and \textit{(3)} termination of the encryption process, are presented. Although the creation of files adds to the resource consumption, the execution of the ransomware is limited to 84 s. Furthermore, due to the creation of dummy files, only 7.1 MB of data are encrypted until the malware is detected.

\section{Summary and Findings}
\label{sec:conclusion}

This work presents four IoT MTD mechanisms focused on \textit{i)} generating dummy files to trap encrypting processes, \textit{ii)} manipulating file extensions to reduce data leakages, \textit{iii)} sanitizing libraries to inhibit actions hiding, and \textit{iv)} shuffling IP addresses to avoid remote control. These four MTD mechanisms are selected and deployed by a proposed hybrid IoT-oriented framework. The framework presents a modular architecture that uses \textit{i)} ML-based behavior classification or predefined rules to decide WHEN MTD mechanisms should be applied, and \textit{ii)} a rule-based module to decide HOW and WHAT MTD mechanisms are deployed. The framework has been deployed on a real IoT spectrum sensor, where its effectiveness and efficiency were evaluated with different multi-purpose malware showing behaviors of botnets, ransomware, backdoors, and rootkits. The ML-based Reactive process achieved an average of 0.98 F1-Score using Random Forest and took about 10 s to classify normal and malicious behaviors. Finally, the performance of the implemented MTD techniques has been verified in terms of attack mitigation and resource consumption, stopping all the attacks satisfactorily and with low impact on CPU, RAM, disk, and sensing spectrum services. Thus, the framework and experiments conducted therewith address key challenges of MTD, such as the evaluation of MTD in real IoT platforms and the deployment of MTD in an intelligent and resource efficient way.

As future work, the optimization of the implemented MTD techniques is planned together with the development of new MTD mechanisms against other malware behaviors. Besides, it is planned to add Reinforcement Learning to the MTD Enforcement module, making it fully automated and adaptive.

\section*{Acknowledgments}

This work has been partially supported by \textit{(a)} the Swiss Federal Office for Defense Procurement (armasuisse) with the CyberTracer and RESERVE (CYD-C-2020003) projects and \textit{(b)} the University of Zürich UZH.

\bibliographystyle{IEEEtran}  
\bibliography{references}

\end{document}